\documentstyle[psfig]{prhep97}
\def \be{\begin{equation}}
\def \ee{\end{equation}}
\def \bea{\begin{eqnarray}}
\def \eea{\end{eqnarray}}
\def \ben{\begin{enumerate}}
\def \een{\end{enumerate}}
\def \bem#1{\renewcommand{\thefootnote}{\arabic{footnote}}\footnote{#1}}
\def \bm{\boldmath}

\def \branch{\mathrm {Br}\,}

\def \cp{C\hspace{-0.1em}P}

\def \ea{{\it et al.}}

\def \eq#1{Eq.~(\ref{#1})}

\def \fig#1{Fig.~\ref{#1}}

\def \GeV{{\mathrm{GeV}}}

\def \heff{H_{\mathrm{eff}}}

\def \nnu{\nonumber}

\def \Re{{\mathrm{Re}}\,}
\def \rf{Ref.~\cite}

\newcommand{\TO}[2]{\stackrel {#1}{\hbox to #2pt{\rightarrowfill}}}

\def \cseff {c_7^{\mathrm{eff}}}
\def \ceff {c_9^{\mathrm{eff}}}
\def \c9eff*{c_9^{\mathrm{eff}*}}
\def \a{\alpha}
\def \b{\beta}
\def \g{\gamma}
\def \G{\Gamma}
\def \d{\delta}

\def \l{\lambda}

\def \m{\mu}
\def \n{\nu}

\def \p{\pi}
\def \r{\rho}

\def \S{\Sigma}

\def \mdh{\hat{m}_d}

\def \sh{\hat{s}}

\def\pl#1#2#3{{\it Phys.~Lett.\/}~{\bf B#1} (19#2) #3}
\def\pp{{\it preprint\/} }
\def\prd#1#2#3{{\it Phys.~Rev.\/}~{\bf D#1} (19#2) #3}

\def\rmp#1#2#3{{\it Rev.~Mod.~Phys.\/}~{\bf #1} (19#2) #3}

\renewcommand{\thefootnote}{\fnsymbol{footnote}}
\makeatletter
\let\chapter\hid@chapter
\makeatother
\begin{document}
\thispagestyle{empty}
\begin{flushright}
PITHA 97/44 \\ hep-ph/9711301 \\ November 1997
\end{flushright}
\vspace{0.5cm}
\begin{center}
\LARGE \bf{\bm $\cp$ Violation in Selected $B$ Decays$^*$}
\end{center}
\setcounter{footnote}{0}
\begin{center}
\sc{ F.~Kr\"uger\bem{Electronic address: krueger@physik.rwth-aachen.de} and 
L.\,M.~Sehgal\bem{Electronic address: sehgal@physik.rwth-aachen.de}}
\\ \it{Institut f\"ur Theoretische Physik (E), RWTH Aachen\\
D-52056 Aachen, Germany}
\end{center}
\vspace{0.5cm}
\thispagestyle{empty}
\centerline{\bf{\uppercase{Abstract}}}
\begin{quotation}
We summarize the results of two papers in which we have studied  $\cp$ violation in inclusive and exclusive decays $b\to d\,e^+e^-$. Two $\cp$-violating effects are calculated: the partial rate asymmetry between $b$ and 
$\bar{b}$ decay, and the asymmetry between $e^-$ and $e^+$ spectra for an untagged $B/\bar{B}$ mixture. These asymmetries, combined with the branching ratio, can potentially determine the parameters $(\r, \eta)$ of the unitarity triangle. We also summarize a paper by Browder \ea\  on a possible $\cp$-violating asymmetry in the inclusive reaction $B\to K^-(K^{*-})X$.
\end{quotation}
\vspace{1cm}

\begin{center}
$^*$Presented by L.\,M. Sehgal at the \\[.7ex]
{\it International Europhysics Conference on High Energy 
Physics\\[.7ex]
19-26 August 1997, Jerusalem, Israel} \\[.7ex]
To appear in the Proceedings
\end{center}
%
%
%
\setcounter{page}{0}
\clearpage
\authorrunning{F.\,Kr\"uger and L.M.\,Sehgal}
\titlerunning{{\talknumber}: $\cp$ Violation in Selected $B$ Decays}
\def \talknumber{902} 
\title{{\talknumber}: \bm$\cp$ Violation in Selected $B$ Decays\thanks{Presented by L.M.\,Sehgal. }}
\author{F.\,Kr\"uger\thanks{Supported by the Deutsche Forschungsgemeinschaft (DFG) 
through Grant No. Se 502/4-3.}  and 
L.M.\,Sehgal}
\institute{Institut f\"ur Theoretische Physik (E), RWTH Aachen,\\
D-52056 Aachen, Germany}
\maketitle
\begin{abstract}
We summarize the results of two papers  in which we have studied  $\cp$ violation in inclusive and exclusive decays $b\to d\,e^+e^-$. Two $\cp$-violating effects are calculated: the partial rate asymmetry between $b$ and $\bar{b}$ decay, and the asymmetry between $e^-$ and $e^+$ spectra for an untagged $B/\bar{B}$ mixture. These asymmetries, combined with the branching ratio, can potentially determine the parameters $(\r, \eta)$ of the unitarity triangle. We also summarize a paper by Browder \ea\  on a possible $\cp$-violating asymmetry in the inclusive reaction $B\to K^-(K^{*-})X$.
\end{abstract}
%
%
\section{Decay $b\to d\, e^+e^-$}
The decay $b\to d\, e^+e^-$ invites attention as a testing ground for $\cp$ for the following reason 
\cite{fklms:incl,fklms:exc}. The 
effective Hamiltonian for $b\to q\, e^+e^-$ $(q= d, s)$ calculated on the basis of electroweak box and penguin diagrams 
(see, for example, \rf{buchalla}) has the structure
\bea
\heff &=&\frac{G_F\a}{\sqrt{2}\p}V_{tb}^{\phantom{*}}V_{tq}^*\bigg\{c_9 ( \bar{q}\g_{\m}
P_L b )
\bar{l}\g^{\m}l+ c_{10}( \bar{q}\g_{\m}P_L b )\bar{l}\g^{\m}\g^5 l\nnu\\[.7ex]
&&\quad\quad\quad\quad\quad\mbox{}-2\, \cseff \bar{q}i  \sigma_{\m\n}\frac{q^{\n}}{q^2}(m_b P_R + m_q P_L)b\, \bar{l}\g^{\m} l\bigg\},
\eea
where the coefficients have numerical values $\cseff=-0.315$, $c_9=4.227$, and 
$c_{10} = -4.642$, and $P_{L,R}= (1\mp \gamma_5)/2$. There is, however, a correction to $c_9$ associated with $u\bar{u}$ and $c\bar{c}$ loop contributions generated by the nonleptonic interaction $b\to q\, u\bar{u}$ and $b\to q\, c\bar{c}$:
\be\label{wilson:nine}
\ceff \approx  c_9 +  \left(3 c_1 + c_ 2 \right) 
\Big\{g(m_c, s)+ \lambda_u\left[ g(m_c, s)- g(m_u, s)\right]\Big\}\ ,
\ee
where the loop functions $g(m_c, s)$ and $g(m_u, s)$ have absorptive parts for $s>4 m_c^2$ and $s>4m_u^2$,
and the coefficient $(3 c_1 + c_ 2)\simeq 0.36$. The complex coefficient $\lambda_u = (V_{ub}^{\phantom{*}}V_{uq}^*)/(V_{tb}^{\phantom{*}}V_{tq}^*)$ is of order $\lambda^2$ in the case of $b\to s\, e^+e^-$ $(q = s)$, but of order unity in the case of 
$b\to d\, e^+e^-$ ($q = d$). Notice that ${\mathrm{arg}}[(V_{ub}^{}V_{ud}^*)/(V_{tb}^{}V_{td}^*)] = \b+\g$, where $\b$, $\g$ are the base angles of the unitarity triangle. For this latter reaction, therefore, the Hamiltonian possesses both the weak (CKM) and dynamical (unitarity) phases that are mandatory for observable $\cp$ violation.

The inclusive aspects of the decay $B\to X_d\, e^+e^-$ can be calculated in the parton model approximation. The differential cross section $\D\Gamma/\D s$ is a quadratic function 
of the couplings $\cseff$, $\ceff$ and $c_{10}$. The spectrum of the antiparticle reaction $\bar{b}\to \bar{d}\, e^+e^-$ is obtained by replacing $\l_u$ by $\l_u^*$, giving rise to a partial rate asymmetry
\be
A_{\cp}(s) = \frac{\D\G/\D s -\D\bar{\G}/\D s}{\D\G/\D s +\D\bar{\G}/\D s} = 
\frac{\eta}{(1-\r)^2 + \eta^2}\times (\mbox{kinematical factor})\ .
\ee

\begin{figure}[p]
\vskip -2truein
\centerline{\psfig{figure=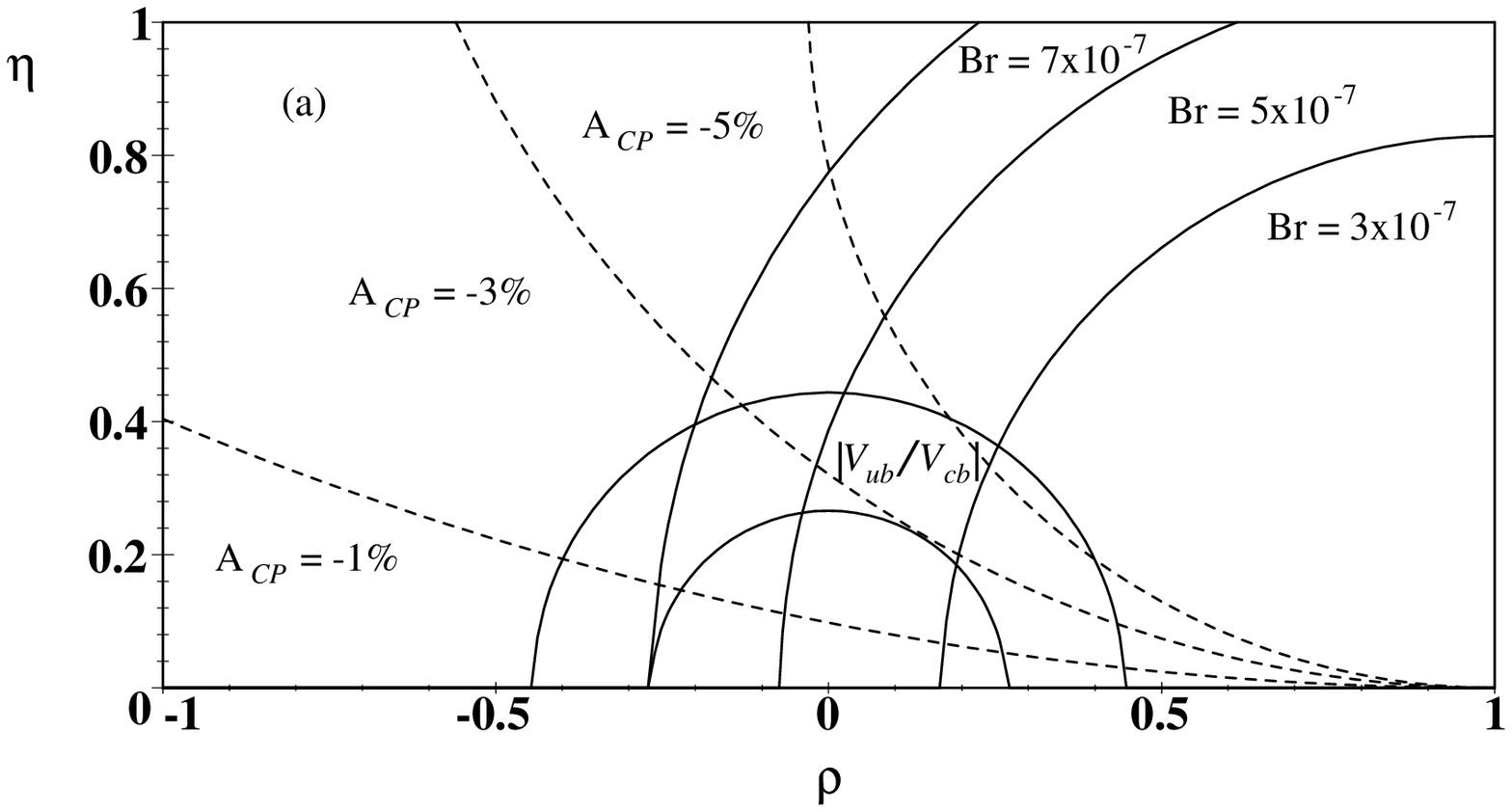,height=5.7in,angle=0}}\vspace{-9.5cm}
\centerline{\psfig{figure=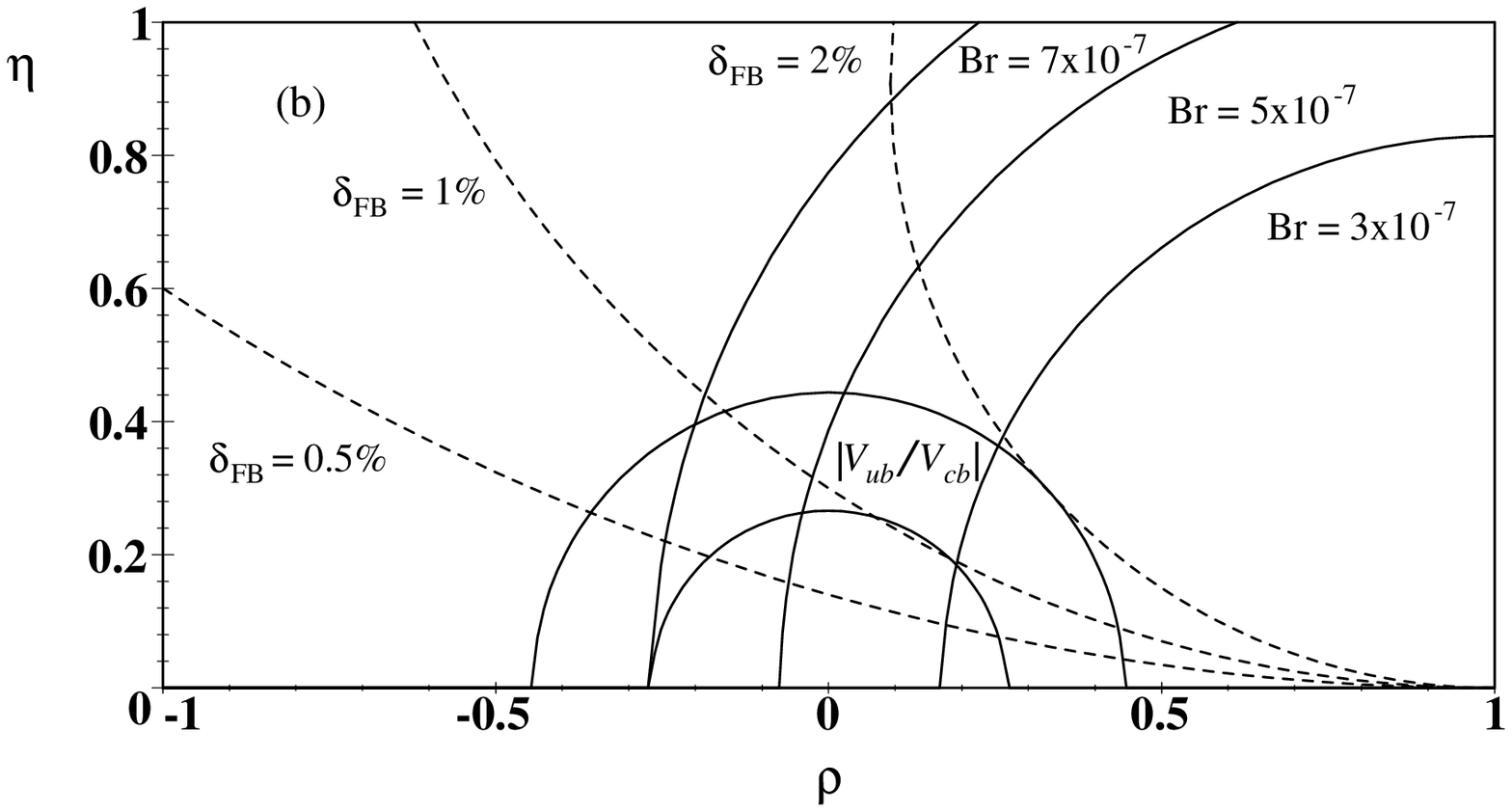,height=5.7in,angle=0}}
\vskip -1.7truein
\caption{Constraints on $(\r, \eta)$ imposed by measurement of branching ratio Br for $B\to X_d\, e^+e^-$, combined with the $\cp$-violating asymmeties  $A_{\cp}$ (a) or $\d_{\mathrm{FB}}$ (b).}\label{fig1talk}
\end{figure}

A second spectral feature is the angular distribution of the $e^-$ in the $e^-e^+$
centre-of-mass system, $\D\G/\D\cos\theta$, which contains a term linear in $\cos\theta$, producing 
a forward-backward asymmetry \cite{alietal} 
\be
A_{\mathrm{FB}}(s)= c_{10}\left[\sh\,\Re  \ceff + 2\cseff\left(1+\mdh^2\right) \right]
\times (\mbox{kinematical factor})
\ee
with the notation $\sh = s/m_b^2$ and  $\mdh = m_d/m_b$.
The corresponding asymmetry of the $e^+$ in $\bar{b}\to \bar{d}\, e^+e^-$ is obtained by replacing
$\l_u$ by $\l_u^*$ in $\ceff$ [cf. \eq{wilson:nine}]. The difference of $A_{\mathrm{FB}}$ and 
$\bar{A}_{\mathrm{FB}}$ is a $\cp$-violating effect
\be
\d_{\mathrm{FB}} = A_{\mathrm{FB}} -\bar{A}_{\mathrm{FB}} = 
c_{10}\frac{\eta}{(1-\r)^2 + \eta^2}\times (\mbox{kinematical factor})\ . 
\ee
Notice that $\d_{\mathrm{FB}} $ is twice the forward-backward asymmetry of the {\it electron} produced by an equal mixture of $B$ and $\bar{B}$ mesons. Equivalently, $\d_{\mathrm{FB}}$ is twice the energy asymmetry of $e^+$ and $e^-$ produced by a $B/\bar{B}$ mixture:
\be
A_{\mathrm E}\equiv \frac{\G(E_+>E_-)-\G(E_+<E_-)}{\G(E_+>E_-)+\G(E_+<E_-)} = 
\frac{1}{2}\d_{\mathrm{FB}} \ ,
\ee
where $E_{\pm}$ denote the lepton energies in the $B$ rest frame.
The branching ratio of  $b\to d\, e^+e^-$ is clearly proportional to $|V_{td}|^2\sim (1-\r)^2 + \eta^2$. A measurement of the branching ratio, together with the asymmetry $A_{\cp}$ or $ \d_{\mathrm{FB}}$
can potentially provide a self-contained determination of the parameters $(\r, \eta)$ of the unitarity triangle, as illustrated in \fig{fig1talk} (for related discussions, see \cite{handokoetal}).
%
%
In \cite{fklms:exc}, we have investigated the exclusive channels $B\to \p\, e^+e^-$ and $B\to \r\, e^+e^-$, employing different models for the form factors \cite{colangelo,melikhov}. Some representative results are given in Table \ref{table1talk}. 
%
%
\begin{table}[p]
\begin{center}
\caption{Branching ratios, forward-backward asymmetries, and the $\cp$-violating asymmetries $A_{\cp}$ and $\d_{\mathrm{FB}}$ in inclusive and exclusive $b\to d\, e^+e^-$ reactions ($\r= -0.07$, $\eta = 0.34$).}\label{table1talk}
\begin{tabular}{ccccc}
\hline\hline\\ [-2.2ex]
& \multicolumn{1}{c}{$B\to X_d\, e^+e^-\quad$}  &
\multicolumn{1}{c}{$B\to\p\, e^+ e^-$ $^{\mathrm b}\quad$}&
\multicolumn{1}{c}{$B\to \r\, e^+ e^-$ $^{\mathrm b}\quad$}
\\[0.4ex] \hline\\ [-2.2ex]
${\branch}\quad$ &$5.5\times 10^{-7}$ & $3.1\times 10^{-8}$ & $5.0\times 10^{-8}$\\ [0.4ex]\hline\\ [-2.2ex]
$\left\langle A_{\mathrm{FB}}\right\rangle\quad$ &$-9\%$&$\equiv 0$ &$-17\%$ \\ [0.4ex]\hline\\ [-2.2ex]
$\left\langle A_{\cp}\right\rangle^{\mathrm a}\quad$ &$-2.7\%$&$-3.1\%$ &$-2.8\%$ \\[0.4ex] \hline\\ [-2.2ex]
$\left\langle\d_{\mathrm{FB}}\right\rangle\quad$& + 1\%& $\equiv 0$ & +2 \%\\[0.4ex]
\hline\hline\\ [-2.2ex]
\multicolumn{4}{l}{$^{\mathrm a}$ for $1\,\GeV< \sqrt{s}< 3\, \GeV$.}\\[0.4ex]
\multicolumn{4}{l}{$^{\mathrm b}$ using form factors of Melikhov and Nikitin \cite{melikhov}.}
\end{tabular}
\end{center}
\end{table}
%
%

\section{$\cp$ Violation in Inclusive $B\to K^{(*)} X$}
In a paper by Browder \ea\ \cite{browderetal}, attention is focussed on the decays 
$B^-\to K^{(*)-} X$ and $\bar{B}^0\to K^{(*)-} X$ with a highly energetic $K$ $(K^*)$, the system $X$ containing 
$u$ and $d$  quarks only. By requiring $E_{K^{(*)}} > 2\, \GeV$, the background from $b\to c\to s$ is effectively suppressed. An attempt is made to relate such ``quasi-inclusive'' decays to the elementary process
$b\to s g^*\to s u\bar{u}$ (QCD penguin). Interference of this process with the tree-level transition $b\to s u\bar{u}$ generates a $\cp$-violating asymmetry between $B\to K^{(*)-} X$ and 
$\bar{B}\to K^{(*)+} X$.

The implementation of this idea involves an effective Hamiltonian containing  a tree-level term 
proportional to $V_{ub}^{}V_{us}^{*}$ and a QCD penguin interaction $\sim (V_{ub}^{}V_{us}^{*}
G_u + V_{cb}^{}V_{cs}^{*}G_c)$, 
where $G_u(q^2)$ and $G_c(q^2)$ are functions denoting the $u\bar{u}$ and $c\bar{c}$ loop contributions to $b\to s g^*$. For the transition $b\to s u\bar{u}$, the essential unitarity phase comes from $G_c(q^2)$ for $q^2> 4m_c^2$. 

The hadronic amplitude for $B\to  K^{(*)-} X$ is simulated by two types of matrix elements, one of which has the kinematical features of $B^-\to K^- u\bar{u}$ (``three-body'' decay) while the other 
resembles $b\to K^- u$ (``two-body'' decay). The latter is the principal source of energetic $K$'s. While the calculation involves several uncertainties, e.g. the choice of $q^2$ in $G_c(q^2)$, asymmetries of order $10\%$, with branching ratios ${\branch} (B\to K X$, $E_K> 2\,\GeV)\sim 10^{-4}$ are shown to be possible. Since evidence for QCD penguins is emerging from the two-body decays of the $B$ meson \cite{neubert}, a search for $b\to sg^*$ in inclusive $B\to K X$, and the associated $\cp$ asymmetry, would be of interest.
%
%
%

\end{document}